\documentclass{article}
\usepackage[utf8]{inputenc} 
\usepackage[T1]{fontenc}    
\usepackage{hyperref}       
\usepackage{url}            
\usepackage{booktabs}       
\usepackage{amsfonts}       
\usepackage{nicefrac}       
\usepackage{microtype}      
\usepackage{graphicx}
\usepackage{caption}

\usepackage{subfig}

\title{Autism Classification Using Brain Functional Connectivity Dynamics and Machine Learning}

\author{
Ravi Tejwani\textsuperscript{1}, Adam Liska\textsuperscript{2},
Hongyuan You\textsuperscript{3}, Jenna Reinen\textsuperscript{1},\\ 
and Payel Das\textsuperscript{1} \\
\textsuperscript{1} IBM Research AI\\
IBM Thomas J. Watson Research Center \\
\texttt{rtejwan@us.ibm.com, jenna.reinen@ibm.com, daspa@us.ibm.com}\\
\textsuperscript{2} CIMeC, Center for Mind/Brain Sciences, University of Trento, Italy \\
Center for Neuroscience and Cognitive Systems @ UniTn,\\
Istituto Italiano di Tecnologia \\
\texttt{adam.liska@gmail.com}\\
\textsuperscript{3} University of California, Santa Barbara \\
\texttt{hyou@cs.ucsb.edu}}

\begin{document}

\maketitle

\begin{abstract}
  The goal of the present study is to identify autism using machine learning techniques and resting-state brain imaging data, leveraging the temporal variability of the functional connections (FC) as the only information. We estimated and compared the FC variability across brain regions between typical, healthy subjects and autistic population by analyzing brain imaging data from a world-wide multi-site database known as ABIDE (Autism Brain Imaging Data Exchange). Our analysis revealed that patients diagnosed with autism spectrum disorder (ASD) show increased FC variability in several brain regions that are associated with low FC variability in the typical brain. We then used the enhanced FC variability of brain regions as features for training machine learning models for ASD classification and achieved 65\% accuracy in identification of ASD versus control subjects within the dataset. We also used node strength estimated from number of functional connections per node averaged over the whole scan as features for ASD classification.The results reveal that the dynamic FC measures outperform or are comparable with the static FC measures in predicting ASD. 
\end{abstract}

\section{Introduction}
There is an increased interest in the dynamics of brain functional connectivity, as measured via temporal variability within the resting-state fMRI (rsfMRI) and task-fMRI BOLD signal, and its aberrations in brain disorders\cite{calhoun2014chronnectome}. Frequently, a sliding-window technique is used, in which functional connections (FC) are estimated within each temporal window lasting on the order of tens of seconds. Novel network measures that characterize dynamic behaviour of individual brain regions have been introduced recently. Node flexibility\cite{bassett2015learning, Bassett_arxiv_2017, shine2016temporal}, which represents the number of times a given brain region (node) switches its functional module assignment from one window to another, has been shown to play a role in learning in a task fMRI paradigm\cite{bassett2015learning}. Another  conceptually simpler measure is node variability\cite{zhang2016neural}, which quantifies the dynamic reconfiguration of the functional connectivity profile of a given node. Characteristic changes in variability have been described in a number of brain disorders, including schizophrenia\cite{braun2016dynamic} and autism spectrum disorders (ASD)\cite{falahpour2016underconnected,zhang2016neural}. However, to what extent the temporal variability of brain regions observed within rsfMRI can be used for neuro-developmental disorder prediction is not extensively studied. 

In this study, we examine the node variability changes in the autistic brain with respect to the typical brain. We further use node variability as features to train machine learning (ML) models for autism identification within a world-wide, large, multi-site functional MRI dataset that consists of both autistic and typical subjects and was collected during resting-state. Our results reveal that supervised machine learning models trained on node variability yield up to 65\% accuracy in ASD classification. Furthermore, we show that equivalent or slightly better prediction accuracy can be obtained by carefully choosing a subset of nodes as features. Finally, node variability yields better or comparable accuracy, when compared to static FC measures, such as node strength, revealing the significance of dynamic FC measures for brain disorder prediction.  

\section{Methods}
The dataset analyzed in the study is the Autism Brain Imaging Data Exchange (ABIDE)\cite{di2014autism}, a large, publicly available dataset of rsfMRI acquisitions of subjects diagnosed with ASD and healthy controls. The pre-processed acquisitions are available as part of the Pre-processed Connectome Project (http://preprocessed-connectomes-project.org/abide/, C-PAC pipeline). We limited ourselves to acquisitions with repetition time of 2s (sites NYU, SDSU, UM, USM). This resulted in a dataset of 147 subjects with ASD and 146 healthy controls. We extracted rsfMRI time series using a 200-region parcellation CC200\cite{craddock2012whole}, estimated subject-wise time-resolved connectivity matrices with windows lengths ranging from 10 to 20 volumes (20-40s) and calculated node variability. The number of subjects and the average mean functional displacement for each site were 119 and 0.0726 for NYU, 94 and 0.0997 for UM, 59 and 0.1097 for USM, and 21 and 0.0696 for SDSU.

The training and testing of machine learning models were performed using the open source machine learning tool kit - Weka\cite{weka_data_minin_softw}. We defined two different sets of features while training the model using variability: (1) all 200 nodes and (2) nodes that show variability <0.9 in typical subjects and experience a positive group-level difference  across the three big sites (NYU, UM, and USM). We also consider training the classifier using node strengths as the features, for which the node strength was defined as a sum of weights (correlations) for all positive connections averaged over the whole fMRI time-series.

We trained the models on these features using Naive Bayes, Random Forest, Support Vector Machines and Multilayer Perceptron algorithm. For intra-site, the evaluation of the model was tested using a 5-fold cross-validation scheme. Two scenario was tested: (1) data from all sites were combined keeping their original proportion intact, and (2)  subjects from each site was proportionately removed to create a more balanced subset. For inter-site, we employed a leave-one-out-site approach that involved training the classifier on all the sites except the one left out for testing. To evaluate each classifier, accuracy, sensitivity, and specificity were estimated.

\section{Results}

\begin{figure}%
    \centering
    \subfloat[Variability difference for time-window = 10]{{\includegraphics[width=5cm]{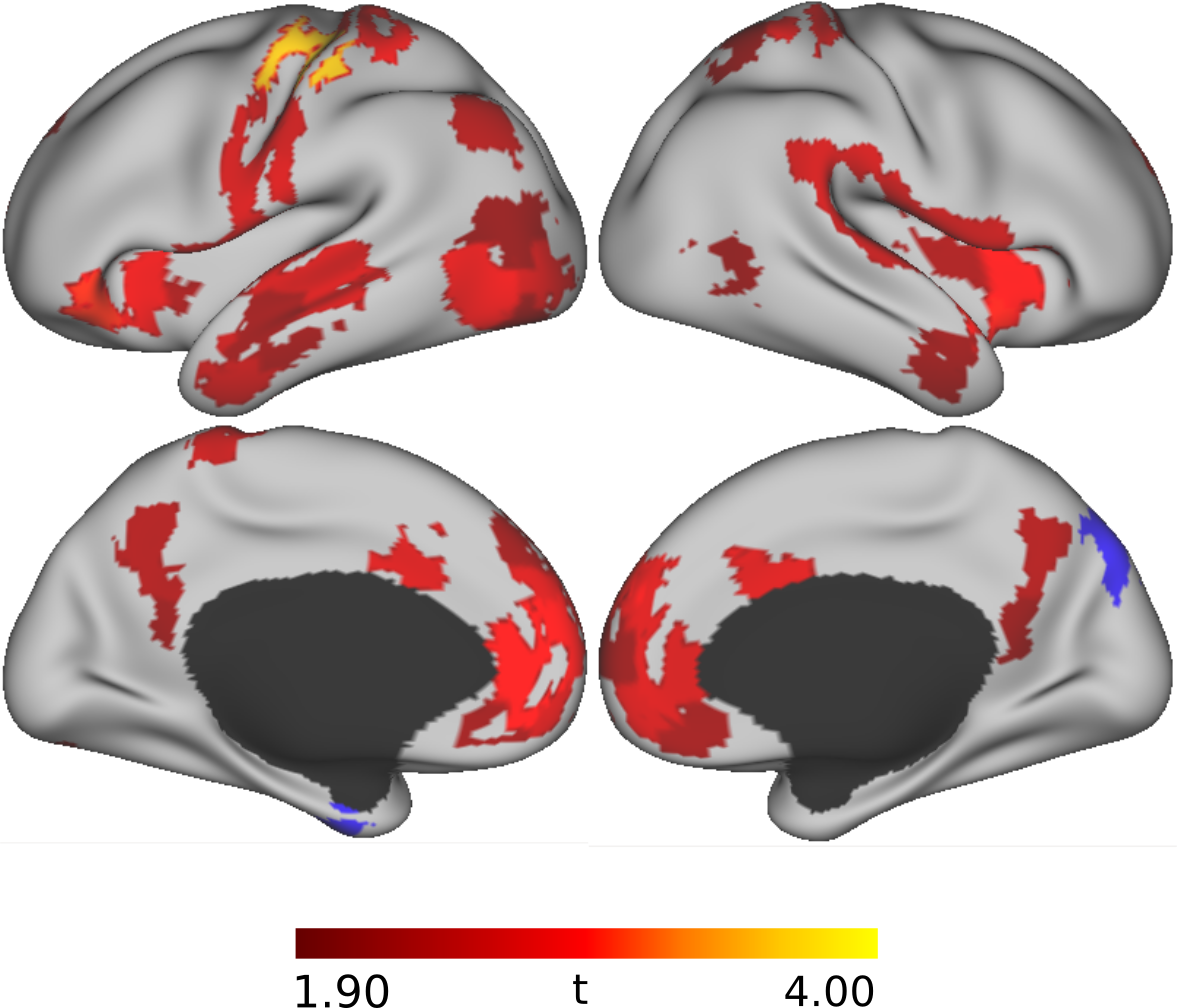} }}%
    \qquad
    \subfloat[Variability difference for time-window = 20]{{\includegraphics[width=5cm]{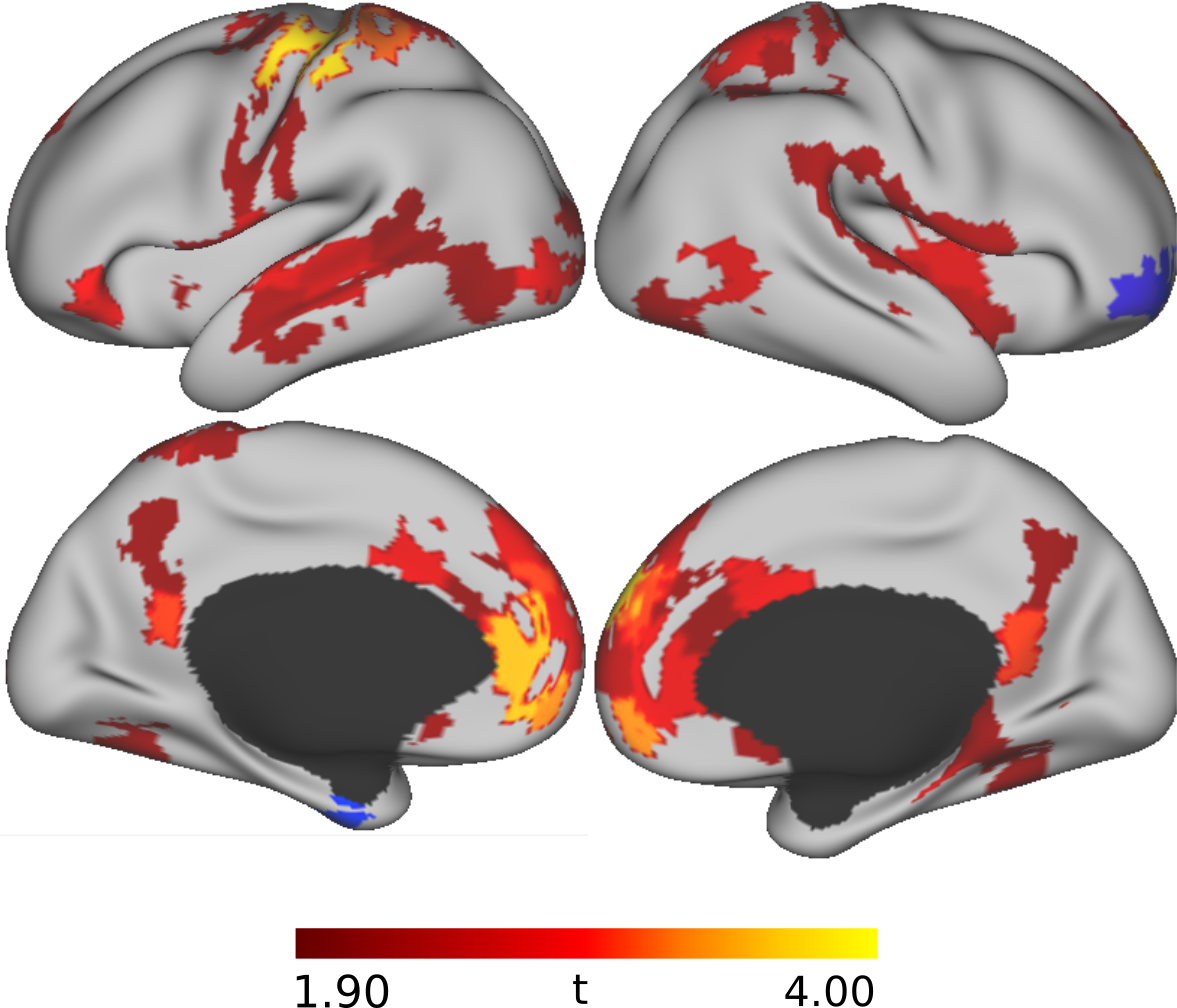} }}%
    \caption{Brain regions showing changes in variability (p-uncorrected < 0.05) are shown in red/yellow (increase) or blue (decrease). The mean
frame-wise displacement measure of each subject was regressed out for the group difference estimation.}%
    \label{fig:example}%
\end{figure}

Figure 1a shows the change in node variability (autistic - typical) across different brain regions, estimated using two different window lengths (20 and 40s). All significant changes represent mostly increase in variability (in red)) in the default and sensorimotor networks within the autistic brain. While the extent of these differences vary with window length, they were present across all window lengths.

\begin{figure}
\centering
\includegraphics[width=10cm]{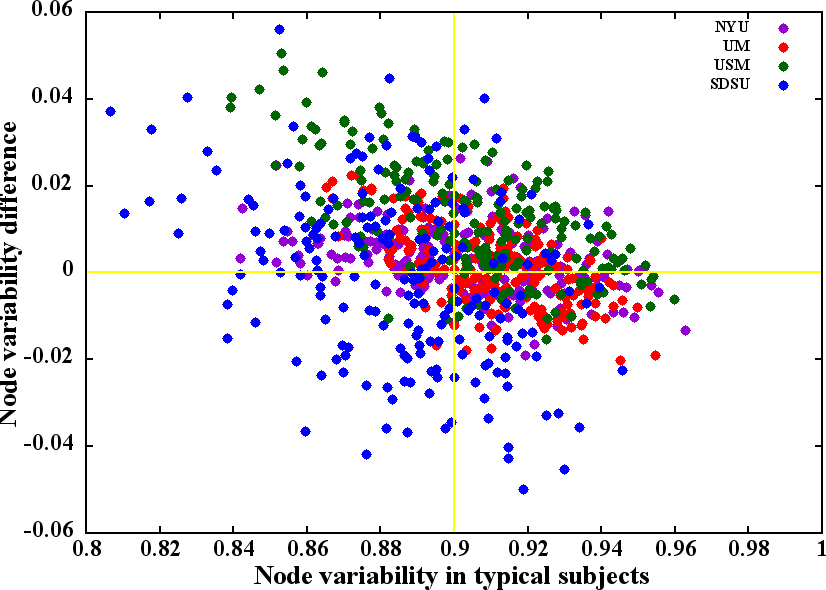}
\caption{Node variability difference and node variability in typical subjects}
\label{fig:output}
\end{figure}

Next, we plotted the node variability difference between two groups as a function of the node variability of the typical subjects for each site. A negative correlation is evident for all sites from Figure \ref{fig:output}, suggesting that low variability nodes experience a variability increase in autism. However, the extent of correlation  between variability difference and variability in typical cohort estimated from Figure 1b appears to be strongly site-dependent (NYU: -0.33; UM: -0.59; USM: -0.64; SDSU: -0.36).

\begin{table}[htp]
\centering
\includegraphics[width=10cm]{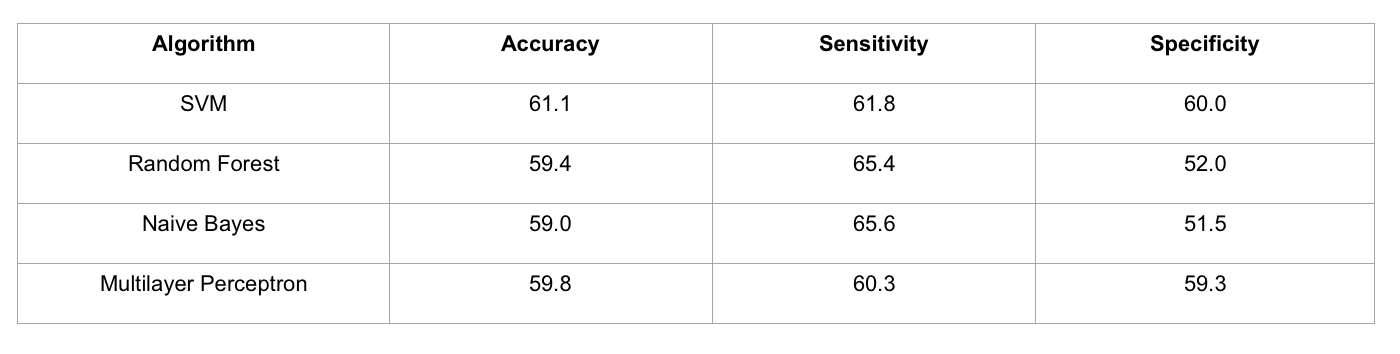}
\caption{Comparison of support vector machine (SVM), Random Forest (RF), Naive Bayes (NB), and Multilayer Perceptron (MLP)
 classifiers trained using 5-fold cross-validation on the entire dataset using node variability as features.}
\label{fig:intrasite}
\end{table}

Table \ref{fig:intrasite} summarizes the performance of different classifiers on intra-site ASD prediction. The accuracy achieved by all classifiers is comparable and is around 60\%. The sensitivity varies between 60 to 65\%. The specificity is \~60\% for SVM and MLP, whereas RF and NB produce slightly lower specificity. The standard deviation is estimated to be around \~3-4\%, as estimated by training and testing the classifier on different sample sizes. Using a subset of low variability nodes that experience variability increase in the ASD cohort as features also results in comparable performance. Furthermore, adding mean frame-wise displacement in the feature set did not affect the performance. 

\begin{table}[htp]
\centering
\includegraphics[width=10cm]{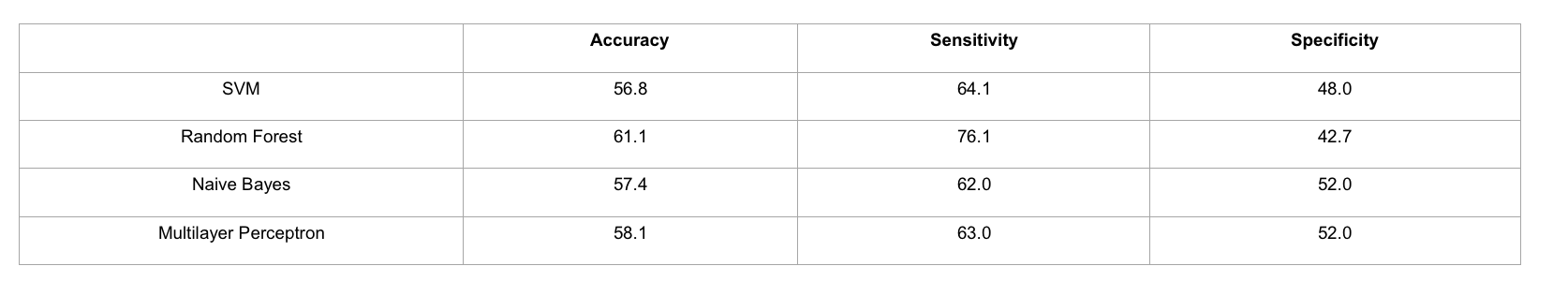}
\caption{Comparison of Support Vector Machine (SVM), Random Forest (RF), Naive Bayes (NB), and Multilayer Perceptron (MLP)
 classifiers trained using 5-fold cross-validation on the entire dataset using node strength as features.}
\label{fig:intrasite-strength}
\end{table}

Table \ref{fig:intrasite-strength} suggests that the models trained on node variability perform slightly better than the models trained  on node strength; suggesting the significance of temporal variability in functional connectome data for brain state/disorder classification. The accuracy obtained using node strength is similar to those reported in prior studies, in which static functional connectivity-based features were used for Autism prediction\cite{heinsfeld2018identification,yahata2016small,resting-state}

\begin{table}[htp]
\centering
\includegraphics[width=10cm]{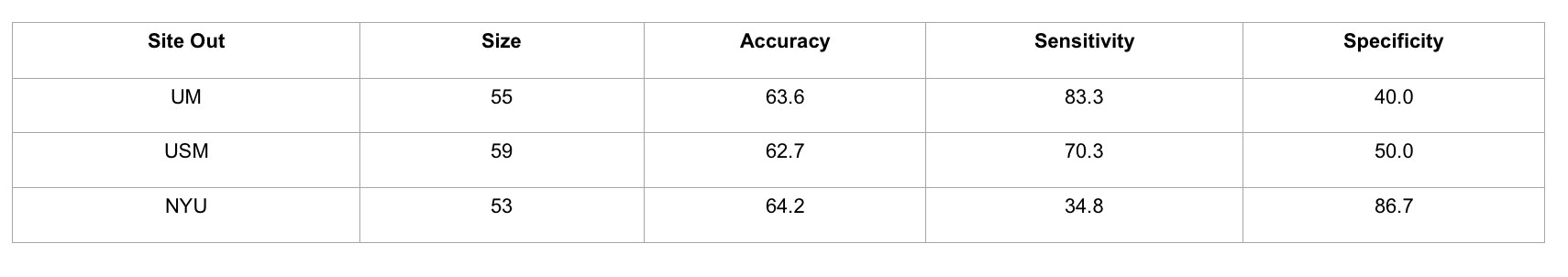}
\caption{Performance of MLP classifier in the leave-one-site-out classification using variability of a subset of nodes as features.}
\label{fig:inter}
\end{table}

The results on inter-site classification are presented in Table \ref{fig:inter}. Whereas the classification accuracy using MLP is \~62\% for all sites, the specificity and  sensitivity vary across sites, consistent with the results shown in Figure 1b. 

\section{Conclusion}
In this study, we used the temporal variability of functional connectivity for ASD classification in a large, multi-site, resting-state fMRI dataset. Our results show that machine learning models trained on brain region variability can yield up to 62\% accuracy, which is comparable with classification accuracy obtained with static connectivity measure such as node strength. In summary, the present study demonstrates the potential of dynamic functional connectivity measures in brain disorder classification.  \newline \newline

\footnotesize{
\bibliographystyle{ieeetr}

\section*{References}
\medskip
\small
[1] V. D. Calhoun, R. Miller, G. Pearlson, and T. Adalı, “The chronnectome: time-varying connectivity networks as the next frontier in fmri data discovery,” Neuron, vol. 84, no. 2, pp. 262–274, 2014. \newline \newline
[2] D. S. Bassett, M. Yang, N. F. Wymbs, and S. T. Grafton, “Learning-induced autonomy of sensorimotor systems,” Nature Neuroscience, vol. 18, no. 5, pp. 744–751, 2015. \newline \newline
[3] P. G. Reddy, M. G. Mattar, A. C. Murphy, N. F. Wymbs, S. T. Grafton, T. D. Satterth- waite, and D. S. Bassett, “Brain state flexibility accompanies motor-skill acquisition,” arXiv preprint arXiv:1701.07646, 2017. \newline \newline
[4] J. M. Shine, O. Koyejo, and R. A. Poldrack, “Temporal metastates are associated with differential patterns of time-resolved connectivity, network topology, and attention,” Proceedings of the National Academy of Sciences, p. 201604898, 2016. \newline \newline
[5] J. Zhang, W. Cheng, Z. Liu, K. Zhang, X. Lei, Y. Yao, B. Becker, Y. Liu, K. M. Kendrick, G. Lu, et al., “Neural, electrophysiological and anatomical basis of brain-network variability and its characteristic changes in mental disorders,” Brain, vol. 139, no. 8, pp. 2307–2321, 2016. \newline \newline
[6] U. Braun, A. Schäfer, D. S. Bassett, F. Rausch, J. I. Schweiger, E. Bilek, S. Erk, N. Romanczuk-Seiferth, O. Grimm, L. S. Geiger, et al., “Dynamic brain network re- configuration as a potential schizophrenia genetic risk mechanism modulated by nmda receptor function,” Proceedings of the National Academy of Sciences, vol. 113, no. 44, pp. 12568–12573, 2016. \newline \newline
[7] M. Falahpour, W. K. Thompson, A. E. Abbott, A. Jahedi, M. E. Mulvey, M. Datko, T. T. Liu, and R.-A. Müller, “Underconnected, but not broken? dynamic functional connectivity mri shows underconnectivity in autism is linked to increased intra-individual variability across time,” Brain Connectivity, vol. 6, no. 5, pp. 403–414, 2016. \newline \newline
[8] A. Di Martino, C.-G. Yan, Q. Li, E. Denio, F. X. Castellanos, K. Alaerts, J. S. Anderson, M. Assaf, S. Y. Bookheimer, M. Dapretto, et al., “The autism brain imaging data exchange: towards a large-scale evaluation of the intrinsic brain architecture in autism,” Molecular Psychiatry, vol. 19, no. 6, pp. 659–667, 2014.  \newline \newline
[9] R. C. Craddock, G. A. James, P. E. Holtzheimer, X. P. Hu, and H. S. Mayberg, “A whole brain fmri atlas generated via spatially constrained spectral clustering,” Human brain mapping, vol. 33, no. 8, pp. 1914–1928, 2012. \newline \newline
[10] M. Hall, E. Frank, G. Holmes, B. Pfahringer, P. Reutemann, and I. H. Witten, “The WEKA data mining software: an update,” SIGKDD Explorations, vol. 11, no. 1, pp. 10–18, 2009. \newline \newline
[11] A.S.Heinsfeld,A.R.Franco,R.C.Craddock,A.Buchweitz,andF.Meneguzzi,“Identifica- tion of autism spectrum disorder using deep learning and the abide dataset,” NeuroImage: Clinical, vol. 17, pp. 16–23, 2018. \newline \newline
[12] N. Yahata, J. Morimoto, R. Hashimoto, G. Lisi, K. Shibata, Y. Kawakubo, H. Kuwabara, M. Kuroda, T. Yamada, F. Megumi, et al., “A small number of abnormal brain connections predicts adult autism spectrum disorder,” Nature communications, vol. 7, 2016. \newline \newline

}

\begin{center}
  \url{https://cmt.research.microsoft.com/NIPS2017/}
\end{center}

\end{document}